\documentclass[aps,prl,twocolumn,epsfig,showpacs,superscriptaddress,footnote]{revtex4}

\usepackage{graphicx}
\usepackage{dcolumn}
\usepackage{bm}

\makeatletter
\usepackage{calc}
\usepackage{times}
\usepackage{multirow}
\def\be{\begin{equation}}
\def\ee{\end{equation}}
\def\bea{\begin{eqnarray}}
\def\eea{\end{eqnarray}}
\def\bma{\begin{mathletters}}
\def\ema{\end{mathletters}}
\def\bi{\begin{itemize}}
\def\ei{\end{itemize}}

\def\C{\hbox{$\mit I$\kern-.7em$\mit C$}}

\newcommand{\singlespacing}{\let\CS=\@currsize\renewcommand{\baselinestretch}
{1.0}\tiny\CS}
\newcommand{\doublespacing}{\let\CS=\@currsize\renewcommand{\baselinestretch}
{1.5}\tiny\CS}

\begin{document}
\title{Complete proof of Gisin's theorem for three qubits}

\author{Sujit K. Choudhary}
\email{sujit@imsc.res.in} \affiliation{The Institute of
Mathematical Sciences, C. I. T. Campus, Taramani, Chennai 600113,
India}

\author{Sibasish Ghosh}
\email{sibasish@imsc.res.in} \affiliation{The Institute of
Mathematical Sciences, C. I. T. Campus, Taramani, Chennai 600113,
India}

\author{Guruprasad Kar}
\email{gkar@isical.ac.in} \affiliation{Physics and Applied
Mathematics Unit, Indian Statistical Institute, 203, B. T. Road,
Kolkata 700 108, India}

\author{Ramij Rahaman}
\email{ramij_r@isical.ac.in} \affiliation{Physics and Applied
Mathematics Unit, Indian Statistical Institute, 203, B. T. Road,
Kolkata 700 108, India}

\begin{abstract}
Gisin's theorem assures that for any pure bipartite entangled state,
there is violation of Bell-CHSH inequality revealing its
contradiction with local realistic model. Whether, similar result
holds for three-qubit pure entangled states, remained unresolved. We
show analytically that all three-qubit pure entangled states violate
a Bell-type inequality, derived on the basis of local realism, by
exploiting the Hardy's non-locality argument.
\end{abstract}



\maketitle Not all measurement correlations in some state of a
composite quantum system can be described by local hidden variable
theory (LHVT) \cite{bell(1964)}, a fact which is said to be ``the
most profound discovery of science'' \cite{stapp(1975)}.
Experimental verification of this fact ({\it i.e.}, whether
measurement correlations in Nature obey quantum rules or LHVT) goes
also in favour of Qunatum Theory, modulo some loopholes
\cite{aspect(2007)}. Every LHVT description \cite{epr-local} of the
measurement correlations of a composite system (assumed to be finite
dimensional in the present paper) gives rise to one (or, more than
one) linear inequality (or, inequalities) involving these
correlations \cite{loubenets(2008)}. There are states of a composite
quantum system, which violate some or all these inequalities for
suitable choices of the subsystem observables.

Gisin's theorem assures that for any pure entangled state of
two-qudits, the above-mentioned violation is generic for two
settings per site ({\it i.e.}, for the choice of one between two
non-commuting observables per qudit) \cite{gisin(1992)}. In other
words, {\it all} pure entangled states of two $d$-dimensional
quantum systems violate a {\it single} Bell-type inequality with two
settings per site, where the choice of the observables depends on
that of the state. However, validity of this theorem for
multi-partite systems is still not guaranteed. For example, for odd
$N$, there is a family of entangled pure states of $N$ qubits, each
of which satisfies {\it all} Bell-type inequalities involving
correlation functions, arising out of measurement of one between two
non-commuting dichotomic observables per qubit
\cite{zukowski(2002)}. Later, Chen et al. \cite{chen(2004)} provided
a Bell-type inequality involving joint probabilities, associated to
measurement of one between two non-commuting dichotomic observables
per qubit, which is violated by all the states of the
above-mentioned family \cite{obschoice}. But a single Bell-type
inequality is not guaranteed to be violated by {\it all} pure
entangled states of three-qubits, although there are claims of
having numerical evidences in favour of this violation
\cite{chen(2004), wu(2008), chen(2008)}.

Quantum Theory also shows contradiction with LHVT via `non-locality
without inequalities' (NLWI) \cite{hardy(1993)}. In this case, a set
of values of joint probabilities of outcomes of measurements of one
between two non-commuting observables per site, has contradiction
with LHVT but can be realized in Quantum Theory. Unfortunately, NLWI
is weaker than Bell-type inequalities, as {\it no} maximally
entangled state of two-qudits seems to show NLWI (in the case of
Hardy-type NLWI, this has been shown in \cite{kaushik(2008)}) even
though each of them violates a Bell-type inequality. This situation
changes drastically when we consider Hardy-type NLWI for three
two-level systems \cite{wu(1996), kar(1997)}, where all but one of
the joint probabilities in the above-mentioned set are zero. {\it
Every} maximally entangled state of three qubits \cite{mes3q}
satisfies Hardy-type NLWI for suitably chosen pairs of non-commuting
dichotomic observables per qubit \cite{ghosh(1998)}. Moreover, an
attempt was made in ref. \cite{ghosh(1998)} towards achieving the
result that {\it every} pure state of three qubits, having genuine
tri-partite entanglement, satisfies Hardy-type NLWI. But in absence
of the discovery of canonical form for three-qubit pure states
(which was done later in ref. \cite{acin(2000)}), it did not yield a
complete proof.

Now, from the set of joint probabilities in any NLWI argument, one
can, in principle, construct a linear inequality involving these
joint probabilities by using local realistic assumption. This
inequality is automatically violated by every quantum state which
satisfies the corresponding NLWI argument. In the case of Hardy-type
NLWI argument for two two-level systems, this inequality (given in
equation (11) of ref. \cite{mermin(1994)}, equation (11) of ref.
\cite{cabello(2002)}, and equation (26) of ref.
\cite{ghirardi(2008)}) is nothing but the corresponding CH
inequality \cite{roy(2007)}. So, by Gisin's theorem, {\it every}
two-qubit pure entangled state (irrespective of its amount of
entanglement) will violate the former inequality. In this letter, we
show {\it analytically} that {\it every} three-qubit pure entangled
state violates a linear inequality of the above-mentioned type (see
eqn. (\ref{hardytobell3q}) below) involving joint probabilities
associated with the Hardy-type NLWI, irrespective of whether the
state has genuine tripartite entanglement or pure bi-partite
entanglement.

Hardy-type NLWI argument starts from the following set of five
joint probability conditions for three two-level systems:
\begin{equation}
\label{3qhardy}
\begin{array}{lcl}
P(D_1 = + 1, U_2 = + 1, U_3 = + 1) &=& 0,\\
P(U_1 = + 1, D_2 = + 1, U_3 = + 1) &=& 0,\\
P(U_1 = + 1, U_2 = + 1, D_3 = + 1) &=& 0,\\
P(D_1 = - 1, D_2 = - 1, D_3 = - 1) &=& 0,\\
P(U_1 = + 1, U_2 = + 1, U_3 = + 1) &>& 0,
\end{array}
\end{equation}
where each $U_j$ (as well as $D_j$) is a $\{+ 1, - 1\}$-valued
random variable. This set of conditions can not be satisfied by a
local realistic theory, and hence it has contradiction with LHVT
\cite{wu(1996), ghosh(1998)}.

To show that in quantum theory there are states which exhibit this
kind of non-locality, we  replace $U_j$ and $D_j$ by the $\{+ 1, -
1\}$-valued observables $\hat{U}_j$ and $\hat{D}_j$ respectively
with $[\hat{U}_j, \hat{D}_j] \neq 0$. The probabilities appearing in
(1) are expectation values of one dimensional projectors
corresponding to the following five product vectors:
\begin{eqnarray*}
|\hat{D}_1 = + 1\rangle |\hat{U}_2 = + 1\rangle |\hat{U}_3 = +
1\rangle,\\
|\hat{U}_1 = + 1\rangle |\hat{D}_2 = + 1\rangle |\hat{U}_3 = +
1\rangle,\\
|\hat{U}_1 = + 1\rangle |\hat{U}_2 = + 1\rangle |\hat{D}_3 = +
1\rangle,\\
|\hat{D}_1 = - 1\rangle |\hat{D}_2 = - 1\rangle |\hat{D}_3 = -
1\rangle,\\
|\hat{U}_1 = + 1\rangle |\hat{U}_2 = + 1\rangle |\hat{U}_3 = +
1\rangle.\\
\end{eqnarray*}
One can easily check that these five vectors are linearly
independent and hence span five dimensional subspace of the eight
dimensional Hilbert space associated to the total system. Hence one
can choose any one (among infinitely many) vector which is
orthogonal to the first four vectors and non-orthogonal to the last
one. Actually this result shows that for any choice of observables
in the above-mentioned non-commuting fashion, one can always find a
quantum state which exhibits contradiction with local realism
\cite{choudhary(2008)}.

But, in this letter, our purpose is to find the converse. We would
like to see whether every three-qubit pure entangled state exhibits
contradiction with local realism. In this context, it has to be
mentioned that Gisin's theorem could provide a single prescription
for finding the observables for any bipartite pure state to show
violation of the Bell-CHSH inequality, due to the existence of
Schmidt decomposition. Schmidt decomposition, in its strict sense
\cite{schmidt}, is absent for systems comprising of three and more
subsystems. This gives rise to complications and one needs to find
the observables for each inequivalent case (depending upon the
values of the parameters describing the state) separately. In this
direction, we start with an arbitrary three-qubit pure state
$|\psi\rangle$, which can always be taken in the canonical form
\cite{acin(2000)}:
\begin{equation}
\label{canonicalform} |\psi\rangle = {\lambda}_0|000\rangle +
{\lambda}_1e^{i{\phi}}|100\rangle + {\lambda}_2|101\rangle +
{\lambda}_3|110\rangle + {\lambda}_4|111\rangle,
\end{equation}
where $0 \le {\lambda}_j$ (for $j = 0, 1, 2, 3, 4$), $\sum_{j =
0}^{4} {\lambda}_j^2 = 1$, and $0 \le \phi \le \pi$.\\
We now fully classify the above-mentioned three-qubit state
$|\psi\rangle$ into four major classes: (A) $|\psi\rangle$ is a
fully product state, (B) $|\psi\rangle$ has pure two-qubit
non-maximal entanglement, (C) $|\psi\rangle$ has pure two-qubit
maximal entanglement, and (D) $|\psi\rangle$ has genuine pure
three-qubit entanglement. Depending on the values of $\lambda_i's$
and $\phi$, in table~\ref{tab:table1}, we further classify each of
these four classes into several sub-classes: (A) consists of (A.1) -
(A.3); (B) consists of (B.1) - (B.5); (C) consists of (C.1) - (C.3);
(D) consists of (D.1) - (D.14).

\newlength{\rcollength}\setlength{\rcollength}{1.3in}
\begin{table}
\caption{\label{tab:table1}Classification of $|\psi\rangle$.}
\begin{tabular}{|l|c|}
 \hline
~~~~~~~Condition & case  \\ \hline

${\lambda}_0{\lambda}_1 \neq 0,~ {\lambda}_2 = {\lambda}_3
={\lambda}_4 = 0$ & (A.1)
\\\hline

 ${\lambda}_0 \ne 0$, ${\lambda}_1 =
{\lambda}_2 = {\lambda}_3 = {\lambda}_4 = 0$ &(A.2)
\\\hline

 ${\lambda}_0 = 0$, ${\lambda}_1{\lambda}_4e^{i{\phi}} =
{\lambda}_2{\lambda}_3$&(A.3)  \\
\hline
 ${\lambda}_0{\lambda}_1{\lambda}_2 \ne 0$, ${\lambda}_3 =
{\lambda}_4 = 0$& (B.1)  \\\cline{1-2}

 ${\lambda}_0{\lambda}_1{\lambda}_3 \ne 0$,
${\lambda}_2 = {\lambda}_4 = 0$& (B.2)
 \\\cline{1-2}

 $0 < {\lambda}_0{\lambda}_2 <
1/2$, ${\lambda}_1 = {\lambda}_3 = {\lambda}_4 = 0$ & (B.3)
 \\\cline{1-2}

 $0 <
{\lambda}_0{\lambda}_3 < 1/2$, ${\lambda}_1 = {\lambda}_2 =
{\lambda}_4 = 0$&(B.4) \\\cline{1-2}

$\lambda_0=0$ and ${\sqrt{2}}\left(\begin{array}{cc}
        {\lambda}_1e^{i{\phi}} & {\lambda}_2\\
        {\lambda}_3 & {\lambda}_4
        \end{array}
  \right)$
 & (B.5)
\\
is neither a singular matrix nor a &\\
 unitary matrix&\\\hline

${\lambda}_0{\lambda}_2 = 1/2$, ${\lambda}_1 = {\lambda}_3 =
{\lambda}_4 = 0$&(C.1)
\\\cline{1-2}

 ${\lambda}_0{\lambda}_3 =
1/2$, ${\lambda}_1 = {\lambda}_2 = {\lambda}_4 = 0$ & (C.2)
\\\cline{1-2}

 $\lambda_0=0$ and ${\sqrt{2}}\left(\begin{array}{cc}
        {\lambda}_1e^{i{\phi}} & {\lambda}_2\\
        {\lambda}_3 & {\lambda}_4
        \end{array}
  \right)$ & (C.3)\\
  is a unitary matrix&
  \\\hline

{${\lambda}_0{\lambda}_1{\lambda}_2{\lambda}_3{\lambda}_4 \ne 0$,
$\phi > 0$}& (D.1) \\\cline{1-2}

  ${\lambda}_0{\lambda}_1{\lambda}_2{\lambda}_3{\lambda}_4 \ne 0$,
$\phi = 0$, ${\lambda}_2{\lambda}_3 \ne {\lambda}_1{\lambda}_4$&
(D.2)
\\\cline{1-2}

 ${\lambda}_0{\lambda}_1{\lambda}_2{\lambda}_3{\lambda}_4 \ne
0$, $\phi = 0$, ${\lambda}_2{\lambda}_3 = {\lambda}_1{\lambda}_4$
& (D.3) \\\cline{1-2}

${\lambda}_0{\lambda}_1{\lambda}_2{\lambda}_3 \ne 0$, ${\lambda}_4
= 0$& (D.4)\\\cline{1-2}

 ${\lambda}_0{\lambda}_1{\lambda}_2{\lambda}_4 \ne 0$, ${\lambda}_3
= 0$& (D.5)
\\\cline{1-2}

${\lambda}_0{\lambda}_1{\lambda}_3{\lambda}_4 \ne 0$, ${\lambda}_2
= 0,~ \lambda_0\neq\lambda_4$& (D.6)\\\cline{1-2}

${\lambda}_0{\lambda}_1{\lambda}_3{\lambda}_4 \ne 0$, ${\lambda}_2
= 0,~ \lambda_0=\lambda_4$& (D.7)\\\cline{1-2}

${\lambda}_0{\lambda}_1{\lambda}_4 \ne 0$, ${\lambda}_2 =
{\lambda}_3 = 0$&(D.8)\\\cline{1-2}

${\lambda}_0{\lambda}_3{\lambda}_4 \ne 0$, ${\lambda}_1 =
{\lambda}_2 = 0$&(D.9)\\\cline{1-2}

${\lambda}_0{\lambda}_2{\lambda}_3{\lambda}_4 \ne 0$, ${\lambda}_1
= 0$, ${\lambda}_2 \ne {\lambda}_4$&(D.10)\\\cline{1-2}

 ${\lambda}_0{\lambda}_2{\lambda}_3{\lambda}_4 \ne 0$, ${\lambda}_1
= 0$, ${\lambda}_2 = {\lambda}_4$&(D.11)\\\cline{1-2}

 ${\lambda}_0{\lambda}_2{\lambda}_3 \ne 0$, ${\lambda}_1 =
{\lambda}_4 = 0$&(D.12)\\\cline{1-2}

${\lambda}_0{\lambda}_2{\lambda}_4 \ne 0$, ${\lambda}_1 =
{\lambda}_3 = 0$&(D.13)\\\cline{1-2}

${\lambda}_0{\lambda}_4 \ne 0$, ${\lambda}_1 = {\lambda}_2 =
{\lambda}_3 = 0$&(D.14)\\\cline{1-2}
 \hline

\end{tabular}
\end{table}

If $|\psi\rangle$ has only bi-partite non-maximal
entanglement we then first consider the situation where
$|\psi\rangle= |\eta\rangle \otimes |\chi\rangle$, with
$|\eta\rangle$ being a two-qubit non-maximally entangled state of
the first and the second qubits, while $|\chi\rangle$ is a state of
the third qubit. Hardy \cite{hardy(1993)} has shown that for all
two-qubit non-maximally entangled pure states, one can choose
observables for both the qubits in such a way that the condition of
non-locality without inequality holds. Now in our three-qubit case,
we first choose $|\hat{U}_3 = + 1\rangle =
(1/{\sqrt{2}})(|\chi\rangle + |{\chi}^{\bot}\rangle)$ and
$|\hat{D}_3 = + 1\rangle = |{\chi}^{\bot}\rangle$, where
${\langle}{\chi}^{\bot}|{\chi}{\rangle} = 0$. We can then choose two
pairs of non-commuting dichotomic observables $(\hat{U}_1,
\hat{D}_1)$ and $(\hat{U}_2, \hat{D}_2)$ in such a way that the
state $|\eta\rangle$ satisfies Hardy's NLWI conditions for two
two-level systems corresponding to these observables . This
immediately shows that the state $|\psi\rangle$ satisfies the
Hardy-type NLWI condition (\ref{3qhardy}). As condition
(\ref{3qhardy}) is symmetric with respect to the qubits, we see that
for each of the cases (B.1) - (B.5), the state $|\psi\rangle$ will
satisfy the Hardy-type NLWI argument (\ref{3qhardy}).

Again, let $|\psi\rangle = |\eta\rangle \otimes |\chi\rangle$, where
$|\eta\rangle$ is a two-qubit maximally entangled state of the first
and the second qubits, while $|\chi\rangle$ is a state of the third
qubit. If we now demand $|\psi\rangle$ to satisfy (\ref{3qhardy}),
it will immediately follow that $|\eta\rangle$ must satisfy Hardy's
NLWI conditions for two two-level systems -- an impossibility
\cite{hardy(1993)}. As above, we see that in none of the cases (C.1)
- (C.3), state $|\psi\rangle$ will satisfy the Hardy-type NLWI
argument.

If $|\psi\rangle $ have genuine tri-partite entanglement then to
show that it satisfies the Hardy-type NLWI argument (\ref{3qhardy}),
one can choose the three pairs of $\{+ 1, - 1\}$-valued
non-commutating observables ($\hat{U}_j, \hat{D}_j$) (where $j$ is
associated with $j$-th system ($j = 1, 2, 3$)) as follows:
$$|\hat{U}_j = + 1\rangle = k_j({\alpha}_j|0\rangle + {\beta}_j|1\rangle),~
|\hat{D}_j = + 1\rangle = l_j({\gamma}_j|0\rangle +
{\delta}_j|1\rangle),$$ where $0 < |k_jl_j({\alpha}_j{\gamma}_j^* +
{\beta}_j{\delta}_j^*)|, |k_jl_j({\alpha}_j{\delta}_j -
{\beta}_j{\gamma}_j)| < 1$, $|k_j{\alpha}_j|^2 + |k_j{\beta}_j|^2 =
|l_j{\gamma}_j|^2 + |l_j{\delta}_j|^2 = 1$, and $k_j$'s, $l_j$'s are
the normalization constants (for $j = 1, 2, 3$). The values of
$\alpha_j,~ \beta_j,~\gamma_j,~\delta_j$ are given in
table~\ref{tab:table2} for all the cases (D.1) - (D.14).

\begin{table*}
\caption{\label{tab:table2}Observables for genuine tri-partite pure entanglement.}
\begin{tabular}{|c|p{\textwidth-1.2\rcollength}|}
\hline
~~~~~~~Case  & ~~~~~~~~~~Set of observables for different cases \\ \hline

 (D.1),~(D.2),~(D.4),~(D.5)&
$\alpha_1=\lambda_1,~\beta_1=-\lambda_0e^{i\phi},~\gamma_1=0,~\delta_1=1$;
$\alpha_2=1,~\beta_2=0,~\gamma_2=\lambda_2\lambda_3e^{i\phi}-\lambda_1\lambda_4,~\delta_2=\lambda_1\lambda_2$
;~$\alpha_3=\lambda_2e^{i\phi},
~\beta_3=-\lambda_1,~\gamma_3=1,~\delta_3=0$ \\

\hline
  (D.3)
&
$\alpha_1=0,~\beta_1=1,~\gamma_1=\lambda_0\lambda_1,~\delta_1=(1-\lambda_o^2);~
\alpha_2=\lambda_1\tau-\lambda_3\epsilon,~\beta_2=\lambda_3\tau+\lambda_1\epsilon,~\gamma_2=\lambda_3
 ,~\delta_2=-\lambda_1;~ \alpha_3=\lambda_1+\lambda_2,~\beta_3=\lambda_2-\lambda_1,~\gamma_3=\lambda_2
 ,~\delta_3=-\lambda_1$, where, $\tau=\lambda_0^2\lambda_3(\lambda_1+\lambda_2),
 ~\epsilon=\lambda_0^2\lambda_1(\lambda_1+\lambda_2)
 +(1-\lambda_0^2)$ \\

 \hline
(D.6)&
$\alpha_1=0,~\beta_1=1,~\gamma_1=\lambda_1e^{-i\phi}(\lambda_4^2-\lambda_0^2),~\delta_1=-\lambda_0(1-\lambda_0^2)
;~\alpha_2=\lambda_3(1-\lambda_0^2),~\beta_2=-\lambda_1e^{-i\phi}(1-\lambda_4^2),~\gamma_2=\lambda_3
,~\delta_2=-\lambda_1e^{-i\phi};~
\alpha_3=1,~\beta_3=0,~\gamma_3=\lambda_4(1-\lambda_4^2)
 ,~\delta_3=\lambda_3(\lambda_4^2-\lambda_0^2)$\\

\hline

(D.7)&
$\alpha_1=\lambda_1e^{-i\phi},~\beta_1=-\lambda_0,~\gamma_1=0,~\delta_1=1
;~\alpha_2=\lambda_3,~\beta_2=-\lambda_1e^{-i\phi},~\gamma_2=1
,~\delta_2=0;~ \alpha_3=1,~\beta_3=0,~\gamma_3=\lambda_0
 ,~\delta_3=-\lambda_3$\\

\hline  (D.8) &
$\alpha_1=0,~\beta_1=1,~\gamma_1=\lambda_1e^{-i\phi}(\epsilon+\lambda_4),~\delta_1=-\lambda_o\epsilon$;
$\alpha_2=1,~\beta_2=1,~\gamma_2=\lambda_4 ,~\delta_2=-\epsilon$;
$\alpha_3=\epsilon,~\beta_3=\lambda_1e^{-i\phi},~\gamma_3=\lambda_4
 ,~\delta_3=-\lambda_1e^{-i\phi}$; with $\epsilon$ being a solution of $z^2(1 - {\lambda}_4^2) +
z{\lambda}_4(1 - {\lambda}_0^2) + {\lambda}_4^4 = 0$  \\

 \hline
 (D.9),~(D.10)& $\alpha_1={\lambda}_2({\lambda}_2^2 +
{\lambda}_4^2) + {\lambda}_4(1 -
{\lambda}_0^2),~\beta_1=-\lambda_0\lambda_3\lambda_4,~\gamma_1=1,~\delta_1=0$;
$\alpha_2=1,~\beta_2=1,~\gamma_2=\lambda_4 ,~\delta_2=-\lambda_2$;
$\alpha_3=0,~\beta_3=1,~\gamma_3=\lambda_3\lambda_4
 ,~\delta_3=\lambda_2^2+\lambda_4^2$  \\

 \hline
 (D.11)&
$\alpha_1=0,~\beta_1=1,~\gamma_1={\lambda}_2^2{\lambda}_3,~\delta_1={\lambda}_0({\lambda}_2^2
+ {\lambda}_3^2)$; $\alpha_2={\lambda}_2^2 +
{\lambda}_3^2,~\beta_2=-\lambda_2^2,~\gamma_2=1 ,~\delta_2=0$;
$\alpha_3=1,~\beta_3=0,~\gamma_3=\lambda_3
 ,~\delta_3=\lambda_2$ \\

 \hline  (D.12) &
$\alpha_1=0,~\beta_1=1,~\gamma_1={\delta}{\lambda}_0{\lambda}_2{\lambda}_3,~\delta_1={\lambda}_2^3{\delta}
+ {\lambda}_3^3$; $\alpha_2=1,~\beta_2=1,~\gamma_2=\lambda_3
,~\delta_2=-\lambda_2\delta$;
$\alpha_3=1,~\beta_3=\delta,~\gamma_3=\lambda_2
 ,~\delta_3=-\lambda_3$; with $\delta$
being a solution of $z^2{\lambda}_2^4 + z{\lambda}_2{\lambda}_3 +
{\lambda}_3^4 = 0$ \\

\hline  (D.13) &
$\alpha_1=1,~\beta_1=1,~\gamma_1=\lambda_2,~\delta_1=-{\lambda}_0\epsilon$;
$\alpha_2=1,~\beta_2=0,~\gamma_2=\lambda_4
,~\delta_2=-({\lambda}_0{\epsilon}+\lambda_2)$;
$\alpha_3=\epsilon,~\beta_3=1,~\gamma_3=\lambda_2
 ,~\delta_3=-\lambda_0$; with
${\epsilon}$ being a solution of $z^2{\lambda}_0^4 +
z{\lambda}_0{\lambda}_2({\lambda}_0^2 + {\lambda}_2^2) +
{\lambda}_2^2({\lambda}_2^2 + {\lambda}_4^2) =
0$ \\

 \hline
 (D.14) &
$\alpha_1=1,~\beta_1=1,~\gamma_1=i\lambda_0,~\delta_1=-{\lambda}_4$;
$\alpha_2=1,~\beta_2=1,~\gamma_2=i\lambda_0
,~\delta_2=-{\lambda}_4$ ;
$\alpha_3=\lambda_4^2,~\beta_3=i\lambda_0^2,~\gamma_3=\lambda_4
 ,~\delta_3=-\lambda_0$\\

  \hline
\end{tabular}
\end{table*}

We now derive a linear inequality (mentioned in equation (7) of ref.
\cite{cereceda(2004)} for $n$ qubits) involving the joint
probabilities in equation (\ref{3qhardy}), starting from local
realistic theory. For this, we first assume that all the
experimental probabilities $P(A_k = i_k)$, $P(A_k = i_k, A_l =
i_l)$, $P(A_k = i_k, A_l = i_l, A_m = i_m)$ (with $A_k \in \{U_k,
D_k\}$, and $i_k \in \{+ 1, - 1\}$ for $k, l, m = 1, 2, 3$, and $k
\ne l \ne m$) can be described by a local hidden variable $\omega$,
defined on the probability space $\Omega$ with probability density
${\rho}({\omega})$. For local realistic theory, the probabilities
would satisfy the following conditions:\\
i) $P_\omega(A_k = i_k)$ (for $i_k \in \{+ 1, - 1\}$ with $k = 1,
2,3$) can only take values 1 or 0. \\
ii)$P_{\omega}(A_k = i_k, A_l = i_l) = P_{\omega}(A_k =
i_k)P_{\omega}(A_l = i_l)$, $P_\omega(A_k = i_k, A_l = i_l, A_m =
i_m) = P_\omega(A_k = i_k)P_\omega(A_l = i_l)P_\omega(A_m = i_m)$.

Condition (i) is equivalent to assigning definite values to the
observables. In any LHVT, the experimental probabilities would be
reproduced in the following way:

$~~P(A_k = i_k) = \int_{\Omega}
{\rho}(\omega)d{\omega}P_{\omega}(A_k = i_k),~P(A_k = i_k, A_l =
i_l) = \int_{\Omega} {\rho}(\omega)d{\omega}P_{\omega}(A_k = i_k,
A_l = i_l),~P(A_k = i_k, A_l = i_l, A_m = i_m) = \int_{\Omega}
{\rho}({\omega})d{\omega}P_\omega(A_k = i_k, A_l = i_l, A_m =
i_m),$
where $\int_{\Omega} {\rho}({\omega})d{\omega} =1.$

Now consider the following quantity
\begin{eqnarray*}
B(\omega) &=& P_\omega(D_1 = -1) P_\omega(D_2 = -1) P_\omega(D_3 =
-1) \\
&+& P_\omega(D_1 = +1) P_\omega(U_2 = +1) P_\omega(U_3 = +1)
\\&+& P_\omega(U_1 = +1) P_\omega(D_2 = +1) P_\omega(U_3 = +1)\\
&+& P_\omega(U_1 = +1) P_\omega(U_2 = +1) P_\omega(D_3 = +1) \\&-&
P_\omega(U_1 = +1) P_\omega(U_2 = +1) P_\omega(U_3 = +1).
\end{eqnarray*}

One can easily check that $B(\omega) \geq 0$ for all
$\omega \in \Omega$. Then obviously
$$\int_{\Omega}{\rho}({\omega})d{\omega} B(\omega)
\geq 0,$$
which, in turn, gives rise to the following Bell-type inequality:
\begin{widetext}
\begin{eqnarray}
\label{hardytobell3q}P(D_1 = - 1, D_2 = - 1, D_3 = - 1) + P(D_1 =
+ 1, U_2 = + 1, U_3 = + 1) + P(U_1 = + 1, D_2 = + 1, U_3 = + 1)
\\\nonumber+ P(U_1 = + 1, U_2 = + 1, D_3 = + 1) - P(U_1 = + 1, U_2 = + 1,
U_3 = + 1) \ge 0.
\end{eqnarray}
\end{widetext} Thus we see that every LHVT satisfies the inequality
(\ref{hardytobell3q}).

From our above-mentioned discussion on Hardy-type NLWI, it follows
that {\it every} three-qubit pure state will violate the
inequality (\ref{hardytobell3q}) unless it is a fully product
state or it has pure bi-partite maximal entanglement. We now show
that this inequality is even violated when $|\psi\rangle$ has pure
bi-partite maximal entanglement, although, in this case,
$|\psi\rangle$ does not satisfy the Hardy-type NLWI condition
(\ref{3qhardy}). Without loss of generality, we can take
$|\psi\rangle$ in this case as: $|\psi\rangle =
(1/{\sqrt{2}})(|00\rangle + |11\rangle) \otimes |0\rangle.$
Choose $|\hat{U}_1 = + 1\rangle = ({\sqrt{0.96}}|0\rangle +
0.2|1\rangle)$, $|\hat{D}_1 = + 1\rangle = |0\rangle$, $|\hat{U}_2
= + 1\rangle = (0.2|0\rangle + {\sqrt{0.96}}|1\rangle)$,
$|\hat{D}_2 = + 1\rangle = |1\rangle$, $|\hat{U}_3 = + 1\rangle =
(1/{\sqrt{2}})(|0\rangle + |1\rangle)$ $ |\hat{D}_3 = + 1\rangle =
|1\rangle$. With this choice, $|\psi\rangle$ will violate the
inequality (\ref{hardytobell3q}).

Thus we have established the desired result that {\it
every pure entangled state of three qubits violates the Bell-type
inequality} (\ref{hardytobell3q}).

If a three-qubit state $|\psi\rangle$ violates the inequality
(\ref{hardytobell3q}) maximally corresponding to the set ${\cal
S}(\psi)$ of three pairs of non-commuting observables $(\hat{U}_1,
\hat{D}_1)$, $(\hat{U}_2, \hat{D}_2)$, and $(\hat{U}_3, \hat{D}_3)$,
then the minimum value of the coefficient $v \in [0, 1]$ for which
the state ${\rho}(\psi, v) \equiv v|{\psi}{\rangle}{\langle}{\psi}|
+ ((1 - v)/8)I$ ($I$ being the $8 \times 8$ identity matrix) also
violates the inequality (\ref{hardytobell3q}), is called here as the
`threshold visibility' of the state $|\psi\rangle$. Lower the amount
of threshold visibility, higher the amount of noise the inequality
can sustain. The maximum negative violation of the inequality
(\ref{hardytobell3q}) by the GHZ state is numerically found to be $-
0.175459$ (approx.), and so the threshold visibility $v_{GHZ}^{thr}$
of this state turns out to be $0.68125$ (approx.), which is
approximately same as that found in \cite{chen(2008)}. On the other
hand, the maximum negative violation of the inequality
(\ref{hardytobell3q}) by the W-state $(1/{\sqrt{3}})(|001\rangle +
|010\rangle + |100\rangle)$ is numerically found to be $- 0.192608$
(approx.), and so the threshold visibility $v_{W}^{thr}$ of this
state turns out to be $0.6606676$ (approx.), which is also
approximately same with the value $0.660668$ of $v^{thr}_W$, found
in \cite{chen(2008)}. It is to be noted that so far as the values of
$v_{GHZ}^{thr}$, $v_W^{thr}$ are concerned, although the
probabilistic Bell-type inequality (18) of ref. \cite{chen(2008)}
and the above-mentioned inequality (\ref{hardytobell3q}) provide
approximately the same values, unlike inequality
(\ref{hardytobell3q}), neither inequality (18) of \cite{chen(2008)}
nor any other inequality, mentioned in the literature till date
(see, for example, \cite{chen(2004), wu(2008), chen(2008)}), is
analytically guaranteed to be violated by all pure entangled states
of three qubits. By considering a modified form of the inequality
(\ref{hardytobell3q}) (e.g., inequality (11) of
\cite{cabello(2002)}), one may get a lower value of the threshold
visibility for the states.

One may also try to find similar feature ({\it i.e.}, violation of
Bell-type inequality, derived from Hardy-type NLWI argument, by all
pure entangled states) in the case of $n$-partite quantum systems.

\section{Acknowledgments}
 R.R. acknowledges the support by CSIR, Government of India,
New Delhi. S.G. would like to gratefully acknowledge the discussion
with S. M. Roy, who pointed out the existence of probabilistic
inequalities, like the one given in equation (\ref{hardytobell3q}),
in the case of any NLWI, and also pointed out the equivalence of the
probabilistic inequality associated to NLWI for two two-level
systems and the corresponding CH inequality. Authors would also like
to thank Samir Kunkri for valuable discussion on Gisin's theorem.


\begin{thebibliography}{99}
\bibitem{bell(1964)} J. S. Bell, {\it Physics} {\bf 1} (1964) 195.
\bibitem{stapp(1975)} H. P. Stapp, {\it Nuovo Cimento B} {\bf 29}
(1975) 270.
\bibitem{aspect(2007)} A. Aspect, {\it Nature} {\bf 446} (2007) 866.
\bibitem{epr-local} Which is also assumed to be `EPR local', {\it i.e.},
one subsystem's probability distribution for measurement outcomes is
independent of the choice of the measurement on other subsystems.
\bibitem{loubenets(2008)} E. R. Loubenets, arXiv:0804.4046 [quant-ph].
\bibitem{gisin(1992)} N. Gisin, {\it Phys. Lett. A} {\bf 154} (1991) 201;
N. Gisin and A. Peres, {\it Phys. Lett. A} {\bf 162} (1992)
15.
\bibitem{zukowski(2002)} M. $\dot{{\rm Z}}$ukowski, $\check{{\rm
C}}$. Brukner, W. Laskowski, and M. Wie$\acute{{\rm s}}$niak, {\it
Phys. Rev. Lett.} {\bf 88} (2002) 210402.
\bibitem{chen(2004)} J.-L. Chen, C. Wu, L. C. Kwek, and C. H. Oh,
{\it Phys. Rev. Lett.} {\bf 93} (2004) 140407.
\bibitem{obschoice} Needless to say that the choice of the observables here
depends on that of the state.
\bibitem{wu(2008)} C. Wu, J.-L. Chen, L. C. Kwek, and C. H. Oh, {\it
Phys. Rev. A} {\bf 77} (2008) 062309.
\bibitem{chen(2008)} J.-L. Chen, C. Wu, L. C. Kwek, and C. H. Oh,
{\it Phys. Rev. A} {\bf 78} (2008) 032107.
\bibitem{hardy(1993)} L. Hardy, {\it Phys. Rev. Lett.} {\bf 71}
(1993) 1665; L. Hardy, {\it Phys. Rev. Lett.} {\bf 68} (1992) 2981.
\bibitem{kaushik(2008)} ``Persistence of non-locality for bi-partite
quantum systems'', P. Kaushik and S. Ghosh (in preparation).
\bibitem{wu(1996)} X.-h. Wu, R.-h. Xie, {\it Phys. Lett. A} {\bf
211} (1996) 129.
\bibitem{kar(1997)} G. Kar, {\it Phys. Rev. A} {\bf 56} (1997)
1023; G. Kar, {\it Phys. Lett. A} {\bf 228} (1997) 119.
\bibitem{mes3q} Any pure state of three qubits, which is
locally unitarily connected to the GHZ state
$(1/{\sqrt{2}})(|000\rangle + |111\rangle)$, is taken here as a
maximally entangled state.
\bibitem{ghosh(1998)} S. Ghosh, G. Kar, and D. Sarkar, {\it Phys. Lett. A} {\bf 243}
(1998) 249.
\bibitem{acin(2000)} A. Ac$\acute{{\rm i}}$n, A. Andrianov, L. Costa, E. Jan$\acute{{\rm
e}}$, J. I. Latorre and R. Tarrach, {\it Phys. Rev. Lett.} {\bf
85} (2000) 1560.
\bibitem{mermin(1994)} N. D. Mermin, {\it Am. J. Phys.} {\bf 62}
(1994) 880.
\bibitem{cabello(2002)} A. Cabello, {\it Phys. Rev. A} {\it 65}
(2002) 032108.
\bibitem{ghirardi(2008)} G. Ghirardi, L. Marinatto, {\it Phys. Lett.
A} {\bf 372} (2008) 1982.
\bibitem{roy(2007)} S. M. Roy, D. Atkinson, G. Auberson, G. Mahoux,
and V. Singh, {\it Mod. Phys. Lett. A} {\bf 22} (2007) 1717.
\bibitem{choudhary(2008)} S. K. Choudhary, S. Ghosh, G. Kar, S.
Kunkri, R. Rahaman, and A. Roy, arXiv:0807.4414 [quant-ph].
\bibitem{schmidt} A `Schmidt decomposed' form for a (multi-partite) state $|\psi\rangle \in
{C\!\!\!\!I}^{d_1} \otimes {C\!\!\!\!I}^{d_2} \otimes \ldots \otimes
{C\!\!\!\!I}^{d_N}$ is given by $|\psi\rangle = \sum_{i = 1}^{d}
{\sqrt{{\lambda}_i}}|{\phi}_{1i}\rangle \otimes |{\phi}_{2i}\rangle
\otimes \ldots \otimes |{\phi}_{Ni}\rangle$, where ${\lambda}_i$'s
are non-negative,
${\langle}{\phi}_{ji}|{\phi}_{ji^{\prime}}{\rangle} =
{\delta}_{ii^{\prime}}$ for all $j = 1, 2, \ldots, N$ and for all
$i, i^{\prime} = 1, 2, \ldots, d \equiv~ {\rm min} \{d_1, d_2,
\ldots, d_N\}$.
\bibitem{cereceda(2004)} J. L. Cereceda, {\it Phys. Lett. A} {\bf
327} (2004) 433.



\end{thebibliography}
\end{document}